\documentclass[aps,prl,twocolumn,groupedaddress,superscriptaddress]{revtex4-1}
\usepackage{graphicx}
\usepackage{amssymb}
\usepackage{epstopdf}
\usepackage{amsmath}
\usepackage{longtable}
\usepackage{amsmath}
\usepackage{amsfonts}
\usepackage[colorlinks=true,citecolor=blue,linkcolor=red]{hyperref}
\usepackage{graphicx}
\usepackage{bbold}     
\usepackage[dvipsnames]{xcolor}

\usepackage{array,booktabs,ragged2e}
\usepackage{soul}      
\usepackage{cancel}      
\usepackage{tabularx}     
\usepackage{multirow}
\usepackage{hhline}
\usepackage{titlesec}

\begin{document}

\title{Large Magnetoresistance and Nontrivial Berry Phase in Nb$_3$Sb Crystals with A15 Structure}

\author{Qin Chen}
\affiliation{Department of Physics, Zhejiang University, Hangzhou 310027, China}

\author{Yuxing Zhou}
\affiliation{Department of Physics, Zhejiang University, Hangzhou 310027, China}

\author{Binjie Xu}
\affiliation{Department of Physics, Zhejiang University, Hangzhou 310027, China}

\author{Zhefeng Lou}
\affiliation{Department of Physics, Zhejiang University, Hangzhou 310027, China}

\author{Huancheng Chen}
\affiliation{Department of Physics, Zhejiang University, Hangzhou 310027, China}
\author{Shuijin Chen}
\affiliation{Department of Physics, Zhejiang University, Hangzhou 310027, China}
\author{Chunxiang Wu}
\affiliation{Department of Physics, Zhejiang University, Hangzhou 310027, China}
\author{Jianhua Du}
\affiliation{Department of Applied Physics, China Jiliang University, Hangzhou $310018$, China}
\author{Hangdong Wang}
\affiliation{Department of Physics, Hangzhou Normal University, Hangzhou 310036, China}
\author{Jinhu Yang}
\affiliation{Department of Physics, Hangzhou Normal University, Hangzhou 310036, China}

\author{Minghu Fang}\email{Corresponding author: mhfang@zju.edu.cn}
\affiliation{Department of Physics, Zhejiang University, Hangzhou 310027, China}
\affiliation{Collaborative Innovation Center of Advanced Microstructure, Nanjing University, Nanjing 210093, China}
\date{\today}
\begin{abstract}

 Compounds with the A15 structure have attracted extensive attention due to their superconductivity and nontrivial topological band structure. We have successfully grown Nb$_3$Sb single crystals with a A15 structure and systematically measured the longitudinal resistivity, Hall resistivity and quantum oscillations in magnetization. Similar to other topological trivial/nontrivial semimetals, Nb$_3$Sb exhibits large magnetoresistance (MR) at low temperatures (717$\%$, 2 K and 9 T), unsaturating quadratic field dependence of MR and up-turn behavior in $\rho_{xx}$(\emph{T}) curves under magnetic field, which is considered to result from a perfect hole-electron compensation, as evidenced by the Hall resistivity measurements. The nonzero Berry phase obtained from the de-Hass van Alphen (dHvA) oscillations demonstrates that Nb$_3$Sb is topologically nontrivial. These results indicate that Nb$_{3}$Sb superconductor is also a semimetal with large MR and nontrivial Berry phase, indicating that Nb$_{3}$Sb may be another platform to search for Majorana zero-energy mode.

\end{abstract}
\maketitle
\section{I. INTRODUCTION}
The recent discovery of extremely large magnetoresistance (XMR) up to 10$^{6}\%$ in nonmagnetic semimetals has inspired tremendous interest in understanding its underlying physical mechanisms and exploring its applications in electronics \cite{ali2014large,tafti2016temperature}. Several mechanisms have been proposed to explain the XMR found in the topologically nontrivial or trivial nonmagnetic semimetals. One scenario attributes the observed linear field dependent MR such as in Cd$_{3}$As$_{2}$ \cite{PhysRevLett.113.246402,liang2015ultrahigh} and Na$_{3}$Bi \cite{PhysRevB.85.195320,xiong2015evidence} to nontrivial topology including the linear band dispersion. The classical carrier compensation mechanism was used to explain the non-saturating quadratic dependence of MR such as in WTe$_{2}$ \cite{ali2014large}, lanthanum monopnictides LaPn (Pn = As, Sb, Bi) \cite{PhysRevB.93.235142,PhysRevLett.117.127204,PhysRevB.94.165163,PhysRevB.96.235128}, as well as in VAs$_{2}$ \cite{chen2021magnetoresistance}. An other mechanism argues that open-orbit trajectories of charge carriers driven by Lorentz force under magnetic field as a result of non-closed Fermi surface to be responsible for XMR as discussed by Zhang \emph{et al}. \cite{PhysRevB.99.035142} and materials such as SiP$_{2}$ \cite{PhysRevB.102.115145} and MoO$_{2}$ \cite{PhysRevB.102.165133} appear to support this picture.

Recently, studies on bulk superconductivity in the materials with topologically nontrivial band structure have drawn a great deal of attention due to the possible realization of the Majorana zero-energy mode (MZM). Many compounds with the A15 structure \cite{stewart2015superconductivity} exhibit high temperature superconductivity, such as Nb$_{3}$Al ($T_{c}$ = 18.7 K), V$_{3}$Si ($T_{c}$ = 16.8 K), Nb$_{3}$Ge ($T_{c}$ = 21.8 K). It has also been suggested that the structure symmetry of A15 compounds with spin-orbit coupling (SOC) give rise to a gapped crossing near the Fermi level. For example, recent theoretical calculations revealed that A15 superconductors Ta$_{3}$Sb, Ta$_{3}$Sn, and Ta$_{3}$Pb have nontrivial band topology in the bulk band structures and topological surface states arise near the Fermi level \cite{PhysRevB.99.224506,derunova2019giant}. In particular, Ta$_{3}$Sb hosts an eightfold degenerate Dirac point close to the Fermi level at the high symmetry point \cite{PhysRevB.99.224506}. First-principles calculations have shown that the gapped Dirac crossings in A15 compounds may result in giant spin Berry curvature and correspondingly giant intrinsic spin Hall effect \cite{derunova2019giant}. Nb$_{3}$Sb with the A15 strucuture is also a superconductor but with very low $T_{c}$ = 0.2 K \cite{knapp1976phonon}. Early research had observed de-Hass van Alphen (dHvA) quantum oscillations and Shubnikov-de Hass (SdH) oscillations under 21 T on Nb$_{3}$Sb single crystals \cite{arko1977haas,sellmyer1980shubnikov}. Recent theoretical calculations indicate that Nb$_{3}$Sb may have nontrivial topology in electronic structures. Gao \emph{et al}. \cite{gao2019topological} predicted various topological semimetals including Nb$_{3}$Sb which is considered to host eightfold degenerate fermions. Zhang \emph{et al}. \cite{zhang2019catalogue} suggested that Nb$_{3}$Sb is a high symmetry point semimetal without SOC taken into consideration, while a topological insulator with SOC. Therefore, we attempt to grow Nb$_{3}$Sb single crystals to study its topological nature.

In this work, based on the successful synthesis of Nb$_{3}$Sb single crystals, we performed comprehensive measurements of longitudinal resistivity, Hall resistivity and the quantum oscillations of magnetization. Our results show that Nb$_{3}$Sb has large unsaturated quadratic field dependent MR (717$\%$, 2 K and 9 T) at low temperatures with up-turn behavior similar to other topological nontrivial/trivial semimetals. The Hall resistivity data indicates that Nb$_{3}$Sb is a perfect hole-electron compensated semimetal. The nonzero Berry phase obtained from the dHvA quantum oscillation of magnetization demonstrates that Nb$_{3}$Sb is topologically nontrivial. These results indicate that the Nb$_{3}$Sb may be a promising platform to investigate the relationship between XMR, topology and superconductivity.

\section{II. EXPERIMENTAL METHODS AND CALCULATIONS}

Nb$_{3}$Sb single crystals were grown by a chemical vapor transport method. Stoichiometrical ratio of high purity Nb powders (99.999$\%$) and Sb powders (99.999$\%$) were sealed in an evacuated quartz tube with 10 mg/cm$^3$ iodine as a transport agent, then heated for two weeks in a tube furnace with a temperature gradient of 1220 -1120 K. Silver grey crystals with typical dimensions 1.0 $\times$ 1.0 $\times$ 0.2 mm$^3$ were obtained at the cold end of the tube. The composition was confirmed to be Nb : Sb = 3 : 1 using the energy dispersive x-ray spectrometer (EDXS). The crystal structure was determined by a powder x-ray diffractometer (XRD, PANalytical), created by grinding pieces of crystals [see Fig. 1(b)]. It was confirmed that Nb$_{3}$Sb crystallizes in a cubic structure (space group \emph{P}m$\bar{3}$n, No. 223). The lattice parameters, \emph{a} =  \emph{b} = \emph{c} = 5.26(2) $\rm {\AA}$ were obtained using Rietveld refinement to the XRD data (weighted profile factor R$_{wp}$ = 7.62$\%$, and the goodness-of-fit $\chi$$^2$ = 2.556), as shown in Fig. 1(b). Electrical resistivity ($\rho$$_{xx}$), Hall resistivity ($\rho$$_{xy}$), and magnetization measurements were carried out using a Quantum Design physical property measurement system (PPMS - 9 T) or Quantum Design magnetic property  measurement system (MPMS - 7 T).

 \begin{figure}[htbp]
\centering
\includegraphics[width=8.6cm]{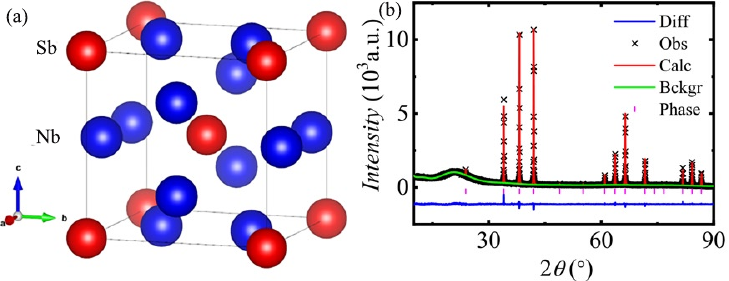}
\caption{(Color online)(a) Crystal structure of the cubic Nb$_3$Sb. (b) XRD pattern of powder obtained by grinding Nb$_3$Sb crystals, the line shows its Rietveld refinement.}
\end{figure}

\section{III. RESULTS AND DISCUSSION}

Firstly, we discuss the longitudinal resistivity $\rho_{xx}$ of Nb$_{3}$Sb measured at various temperatures and in different magnetic fields. The resistivity measured at $\mu_{0}H$ = 0 T [see Fig. 2(a)] exhibits a metallic behavior with $\rho$(2 K) = 0.29 $\mu$$\Omega$~cm, and $\rho$(300 K) = 18 $\mu$$\Omega$~cm, thus the residual resistivity ratio (RRR) $\rho$(300 K)/$\rho$(2 K) $\sim$ 62, indicating that the Nb$_{3}$Sb crystal has a high quality. It was found that Nb$_{3}$Sb crystal exhibits large magnetoresistance. As shown in the inset of Fig. 2(a), an up-turn behavior in $\rho$$_{xx}$(\emph{T}) curves is observed under applied magnetic field at low temperatures: $\rho$$_{xx}$ increases with decreasing \emph{T} and then saturates. Figure 2(b) shows MR as a function of temperature at various magnetic fields, with the conventional definition  $\textit{MR} = \frac{\Delta\rho}{\rho(0)} = [\frac{\rho(H)-\rho(0)}{\rho(0)}]\times100\%$. The normalized MR, shown in the inset of Fig. 2(b), has the same temperature dependence for various fields, excluding the suggestion of a field-induced metal-insulator transition \cite{zhao2015anisotropic,khveshchenko2001magnetic} at low temperatures, as discussed in our work on the topologically trivial semimetal $\alpha$-WP$_2$ \cite{du2018extremely}, and Thoutam \emph{et al}. worked on the type-II Weyl semimetal WTe$_2$ \cite{thoutam2015temperature}.

Figure 3(a) shows the MR as a function of field at different temperatures. The measured MR is large at low temperatures, reaching 717$\%$  at 2 K and 9 T, and having no sign of saturation up to the highest field (9 T) applied in our measurements. In addition, MR can be described by the Kohler scaling law \cite{pippard1989magnetoresistance}
\begin{eqnarray}
\textit{MR} = \frac{\Delta\rho_{xx}(T,H)}{\rho_{0}(T)} = \alpha(\textit{H}/\rho_{0})^{m}.
\end{eqnarray}
As shown in Fig. 3(b), all the MR data measured from $\emph{T}$ = 2 to 100 K collapse onto a single straight line when plotted as MR $\sim $ \emph{H}/$\rho_{0}$ curve, with $\alpha$ = 0.026 ($\mu\Omega$ cm/T)$^{1.8}$ and \emph{m} = 1.8 obtained by fitting. The nearly quadratic field dependence of MR observed for this semimetal Nb$_{3}$Sb is attributed to the perfect electron-hole compensation, as evidenced by the the Hall resistivity measurements discussed below, which is a common characteristic for most of the topologically  nontrivial and trivial semimetals \cite{wang2017large,du2018extremely,chen2018large}.

\begin{figure}[!htbp]
\centering
\includegraphics[width=8.6cm]{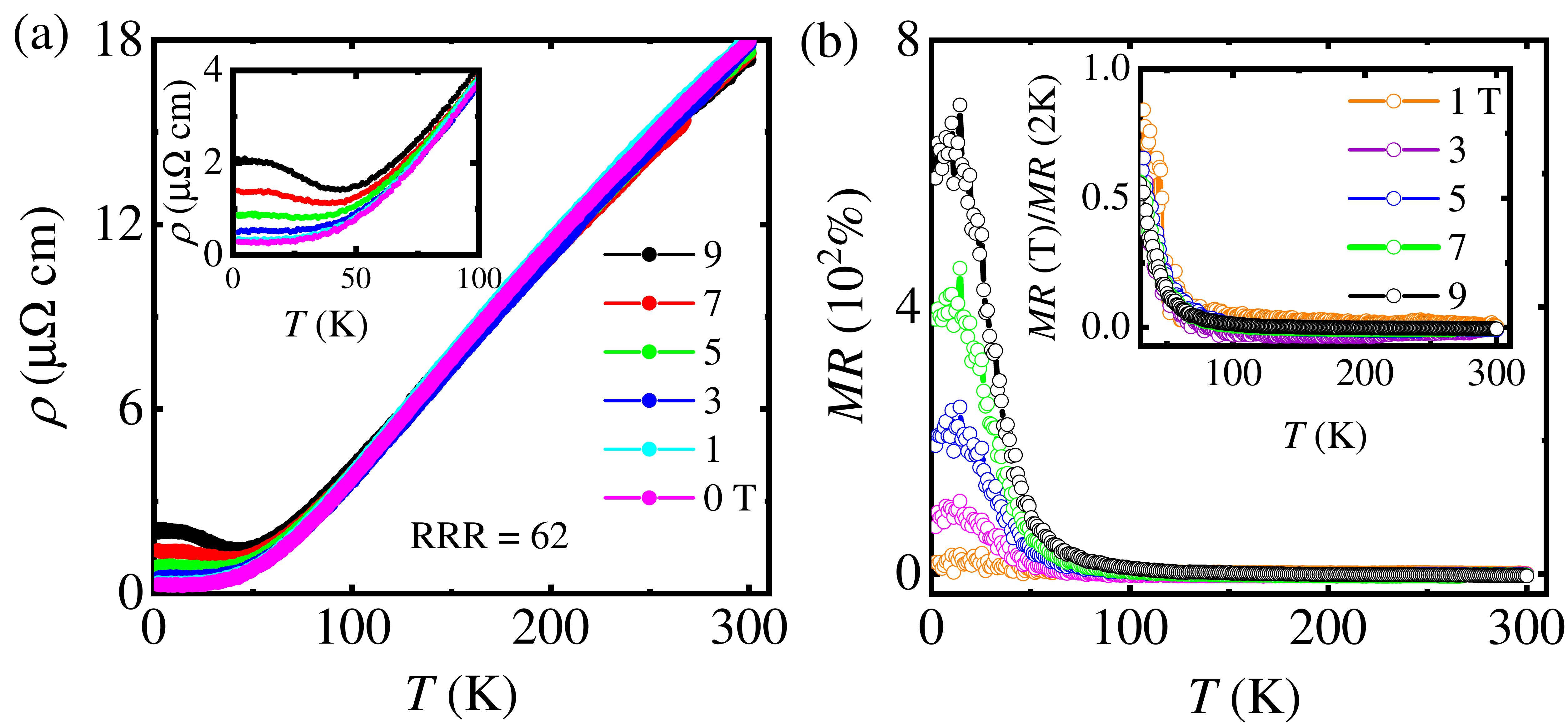}
\caption{(Color online)(a) Temperature dependence of longitudinal resistivity, $\rho$$_{xx}$(\emph{T}), measured at various magnetic fields for a Nb$_3$Sb crystal. The inset shows the low temperature data. (b) The MR as a function of temperature measured  at various magnetic fields.  The inset shows the temperature dependence of MR normalized by its value at 2 K at various magnetic fields.}
\end{figure}

\begin{figure}[!htbp]
\centering
\includegraphics[width=8.6cm]{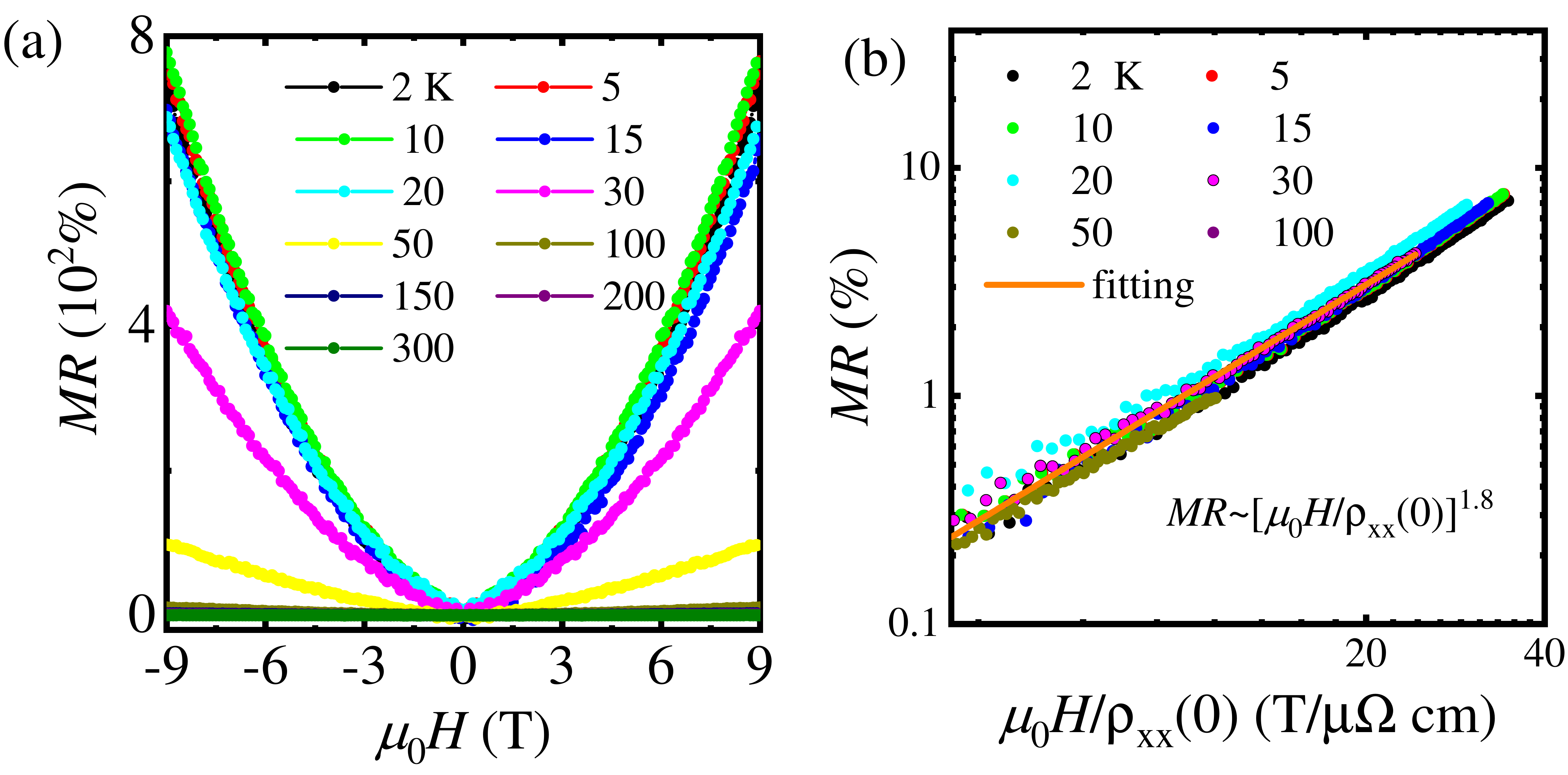}
\caption{(Color online) (a) Field dependence of MR at various temperatures. (b) MR as a function of \emph{H}/$\rho$$_{xx}$(0) plotted on a log scale.}
\end{figure}

\begin{figure}[!htbp]
\centering
\includegraphics[width=8.6cm]{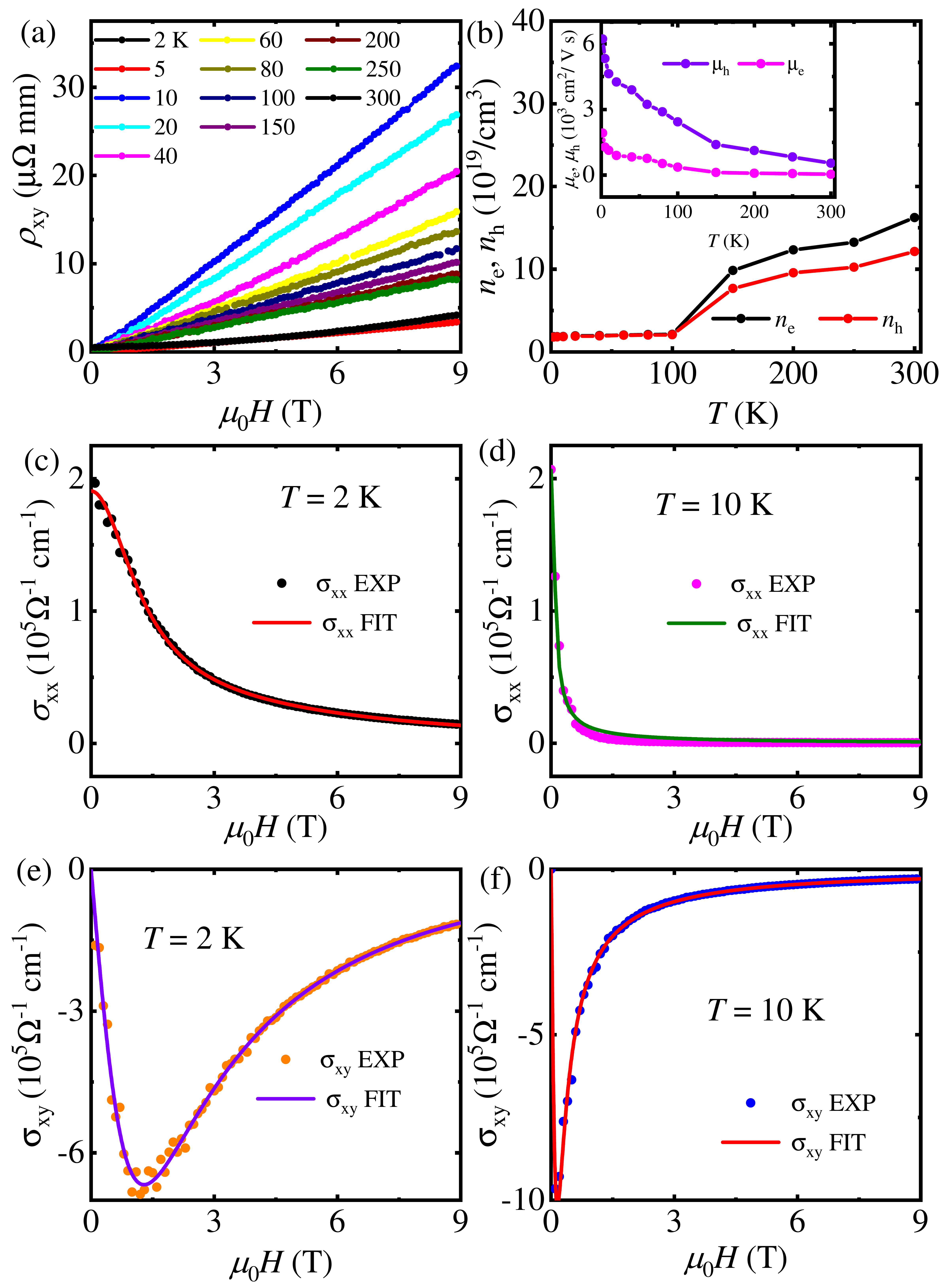}
\caption{(Color online)(a) Field dependence of Hall resistivity $\rho_{xy}$ at different temperatures for a Nb$_3$Sb crystal. (b) The carrier concentrations, \emph{n}$_e$ and  \emph{n}$_h$, and (inset) charge-carrier mobilities, $\mu_e$ and $\mu_h$, as a function of temperature extracted from the two-carrier model. Components of the conductivity tensor, \emph{i}.\emph{e}., $\sigma_{xx}$ and $\sigma_{xy}$ in panels (c), (d), (e) and (f), respectively, as functions of magnetic field for temperatures at 2 K and 10 K. Dots represent experimental data and solid lines the fitting curves based on the two-carrier model.}
\end{figure}

\begin{figure}[!htbp]
\centering
\includegraphics[width=8.6cm]{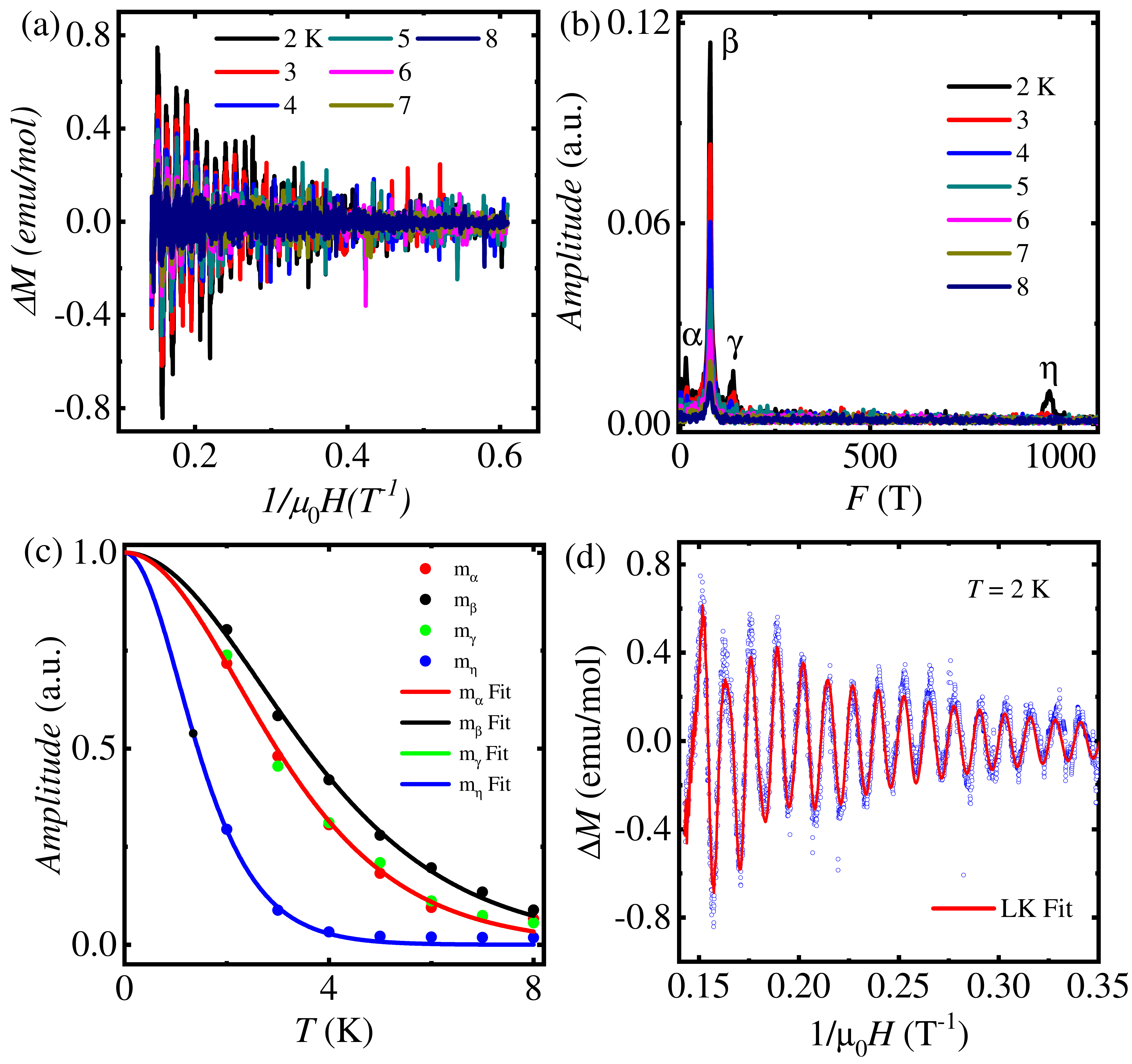}
\caption{(Color online)(a) The amplitude of dHvA plotted as a function of 1/$\mu_0$\emph{H}. (b) FFT spectra of the oscillations between 2 K and 8 K. (c) The temperature dependence of relative FFT amplitude of frequency and the fitting result by \emph{R}$_T$. (d) The fitting of dHvA oscillations at 2 K by the multi-band LK formula.}
\end{figure}

We then discuss the characteristics of charge carriers in Nb$_3$Sb relying on the Hall resistivity measurements. Figure 4(a) displays $\rho_{xy}$(\emph{H}) measured at various temperatures. Although the linear dependence of $\rho_{xy}$(\emph{H}) with a positive slope indicates that the holes dominate the transport properties, it is not true for a semimetal, in which both electron and hole carriers coexist, as discussed by us for the MoO$_2$ \cite{PhysRevB.102.165133}. Following the analysis of $\gamma$-MoTe$_2$ by Zhou \emph{et al.} \cite{zhou2016hall}, as well as for MoO$_2$ by ourself \cite{PhysRevB.102.165133}, we analysed the longitudinal and Hall resistivity data using the semiclassical two-carrier model. In this model, the conductivity tensor, in the complex representation, is given by \cite{ali2014large}
\begin{eqnarray}
 \sigma = \frac{en_{e}\mu_{e}}{1+i\mu_{e}\mu_0H}+\frac{en_{h}\mu_{h}}{1-i\mu_{h}\mu_0H}.
\end{eqnarray}
where \emph{n}$_e$ and \emph{n}$_h$ denote the electron and hole concentrations, $\mu_e$ and $\mu_h$ denote the mobilities of electrons and holes, respectively. From Eq. (2), the Hall conductivity $\sigma_{xy}$ is the imaginary part and the longitudinal conductivity $\sigma_{xx}$ is the real part as shown in the Eq. (3) and (4) \cite{zhou2016hall}.
\begin{eqnarray}
\sigma_{xy} =  \frac{e\mu_0Hn_{h}\mu_{h}^{2}}{1+\mu_{h}^{2}\mu_0^{2}H^{2}}-\frac{e\mu_0Hn_{e}\mu_{e}^{2}}{1+\mu_{e}^{2}\mu_0^{2}H^{2}},
\end{eqnarray}

\begin{eqnarray}
\sigma_{xx} = \frac{en_{h}\mu_{h}}{1+\mu_{h}^{2}\mu_0^{2}H^{2}}+\frac{en_{e}\mu_{e}}{1+\mu_{e}^{2}\mu_0^{2}H^{2}}.
\end{eqnarray}
To appropriately evaluate the carrier densities and mobilities, we calculated the Hall conductivity $\sigma_{xy}$ = $-$ $ \rho_{xy}/(\rho_{xx}^{2}+\rho_{xy}^{2})$ and the longitudinal conductivity $\sigma_{xx}$ = $\rho_{xx}/(\rho_{xx}^{2}+\rho_{xy}^{2})$ using the experimentally observed Hall resistivity $\rho_{xy}$ (\emph{H}) and longitudinal resistivity $\rho_{xx}$ (\emph{H}). Then, we fit both $\sigma_{xy}$(\emph{H}) and $\sigma_{xx}$(\emph{H}) data using Eq. (3) and (4).

Figures 4(c)-(f) display the fits of both the $\sigma_{xx}$ (\emph{H}) and $\sigma_{xy}$ (\emph{H}) at \emph{T} = 2 K and 10 K, respectively. The excellent agreement between our experimental data and the two-carrier model over a broad range of temperatures confirms the coexistence of electrons and holes in Nb$_3$Sb. Figure 4(b) shows the \emph{ n}$_e$, \emph{n}$_h$, $\mu_e$ and $\mu_h$  values obtained by the fitting over the temperature range 2 $-$ 300 K. It is remarkable that the \emph{n}$_e$ and \emph{n}$_h$ values are almost the same below 100 K, such as at 2 K, \emph{n}$_e$ = 1.83 $\times$ 10$^{19}$ cm$^{-3}$, and \emph{n}$_h$ = 1.79 $\times $ 10$^{19}$ cm$^{-3}$. These results indicate that the MR in Nb$_3$Sb semimetal indeed results from the perfect compensation of the two kinds of charge carriers, similar to that observed in many trivial and nontrivial topological semimetals \cite{takatsu2013extremely,mun2012magnetic,yuan2016large,huang2015observation,ali2014large,chen2016extremely}. Due to the existence of phonon thermal scattering at higher temperatures, as shown in the inset of Fig. 4(b), it is obvious that both $\mu_e$ and $\mu_h$ decrease notably with increasing temperature. It is worth noting that hole mobility $\mu_h$ is larger than $\mu_e$ at lower temperatures, such as at 2 K, $\mu_h$ = 6.2 $\times $ 10$^{3}$ cm$^{2}$/V s, and $\mu_e$ = 1.9 $\times $ 10$^{3}$ cm$^{2}$/V s.

Finally, to obtain additional information on the electronic structure, we measured the dHvA quantum oscillations in the isothermal magnetization,\emph{ M}(\emph{H}), for a Nb$_3$Sb crystal up to 7 T for the \emph{H} $\parallel$ \emph{c} axis orientation. After subtracting a smooth background from the \emph{M}(\emph{H}) data at each temperature, the periodic oscillations are visible in 1/\emph{H}, as shown in Fig. 5(a). From the fast Fourier transformation (FFT) analysis, we derived four basic frequencies [Fig. 5(b)]. In general, the oscillatory magnetization of three dimensional (3D) system can be described by the Lifshitz-Kosevich (LK) formula \cite{lifshitz1956theory,shoenberg1984magnetic} with the Berry phase \cite{mikitik1999manifestation}:

\begin{eqnarray}
\Delta M \varpropto -B^{1/2}R_{T}R_{D}R_{S}sin[2\pi(\emph{F}/\emph{B}-\gamma-\delta)],
  \end{eqnarray}
 \begin{eqnarray}
  R_{T}=\alpha T\mu/B sinh(\alpha T\mu/B),
  \end{eqnarray}
  \begin{eqnarray}
    R_{D}=exp(-\alpha T_{D}\mu/B),
  \end{eqnarray}
  \begin{eqnarray}
  R_{S}=cos(\pi g\mu/2).
   \end{eqnarray}

  \begin{table}[htbp]
  \renewcommand\arraystretch{1.2}
  \centering
  \caption{The obtained parameters by fitting dHvA data for Nb$_{3}$Sb}\label{1}
  \setlength{\tabcolsep}{4.0mm}
  {
  \begin{center}
  \begin{tabular}{ccccc}
    \toprule
    Parameters &F$_{\alpha}$&F$_{\beta}$&F$_{\gamma}$ &F$_{\eta}$\\

    \hline
    Frequency (T)& 15 & 79 & 139 & 970 \\
    \textit{m}$^{*}$/\textit{m}$_{0}$ & 0.21 & 0.17& 0.21 & 0.41 \\
    \emph{T}$_{D}$ (K)& 5.2 & 8.0&4.1&9.0 \\
    $\tau_{q}$ (ps)& 0.23 &0.15 &0.30 &0.13\\
    $\mu_{q}$ (cm$^{2}$/Vs)& 1925 & 1511 &2512&5574 \\
    $\phi_{B}$($\delta$=+1/8) &0.65$\pi$&0.89$\pi$&0.19$\pi$&0.49$\pi$\\
    $\phi_{B}$($\delta$=$-$1/8) &0.15$\pi$&0.39$\pi$&0.69$\pi$&1.99$\pi$\\

    \botrule

  \end{tabular}
   \end{center}
  }
\end{table}

where $\mu$ is the ratio of effective cyclotron mass \emph{m}$^*$ to free electron mass \emph{m}$_0$. \emph{T}$_\emph{D}$ is the Dingle temperature, and $\alpha$ = ($2\pi^2 $$\emph{k}_\emph{B} \emph{m}_0$)/($\hbar$e). The phase factor $\delta$ = 1/8 or $-$ 1/8 for three dimensional system. The effective mass \emph{m}$^*$ can be obtained by fitting the temperature dependence of the oscillation amplitude \emph{R}$_\emph{T}$(T), as shown in Fig. 5(c). For \emph{$F_\beta$} = 79 T, the obtained \emph{m}$^*$ is 0.17\emph{m}$_0$. Using the fitted \emph{m}$^*$ as a known parameter, we can further fit the oscillation patterns at a given temperatures [e.g., \emph{T} = 2 K, see Fig. 5(d)] to the LK formula with the frequency, from which quantum mobility and the Berry phase $\phi_\emph{B}$ can be extracted. The fitted Dingle temperature \emph{T}$_\emph{D}$ is of 8.0 K, which corresponds to the quantum relaxation time $\tau_q$ = $\hbar$/(2$\pi$$\emph{$k_BT_D$}$) being of 0.15 ps, and quantum mobility $\mu_q$ = e$\tau$/\emph{m}$^*$ being of 1551 cm$^2$/V s. The LK fit also yields a phase factor $-$ $\gamma$$-$ $\delta$ of 0.82, in which $\gamma$ = $\frac{1}{2}$ $-$ $\frac{\phi_\emph{B}}{2\pi}$ and the Berry phase $\phi_\emph{B}$ is determined to be 0.89$\pi$ for $\delta$ = 1/8 and 0.39$\pi$ for $\delta$ = $-$ 1/8. As is well known, topological nontrivial materials requires a nontrivial $\pi$ Berry phase, while for the trivial materials, the Berry phase equals 0 or 2$\pi$ . In our sample, the Berry phase is close to the $\pi$, so it is a topological nontrivial materials. All the results for other frequencies are listed in Table I.
\section{IV. SUMMARY}
In summary, Nb$_3$Sb with the A15 structure have been studied in detail by longitudinal resistivity, Hall resistivity, and dHvA oscillations in magnetization measurements. It is found that the MR exhibits a field induced up-turn behavior with unsaturated \emph{H}$^{1.8}$ dependence, which is believed to arise from the carrier compensation, as evidenced by the Hall resistivity results. We also obtained the non-zero Berry phase from the dHvA oscillations, indicating the nontrivial topology of band structure in Nb$_3$Sb. Therefore, it is vital to study the topological nature of Nb$_3$Sb in superconducting state in future studies.

\section{\textbf{ACKNOWLEDGMENTS}}

This work was supported by the National Key Program of China (Grant No. 2016YFA0300402), the National Natural Science Foundation of China (Grant Nos. 12074335, and 11974095), the Fundamental Research Funds for the Central Universities.
\bibliography{document}
\end{document}